\documentclass[twocolumn,preprintnumbers,amsmath,amssymb,final]{revtex4}

\usepackage{graphicx}
\usepackage{graphics}
\usepackage{dcolumn}
\usepackage{bm}

\def\BEq{\begin{equation}}
\def\EEq{\end{equation}}
\def\BEqA{\begin{eqnarray}}
\def\EEqA{\end{eqnarray}}
\def\BEn{\begin{enumerate}}
\def\EEn{\end{enumerate}}
\def\BWT{\begin{widetext}}
\def\EWT{\end{widetext}}
\def\a{\alpha}

\def\d{\delta}

\def\g{\gamma}

\begin{document}


\title{Derivation of the Lorentz transformation without the use of Einstein's second postulate}

\author{Andrei Galiautdinov}
 \affiliation{Department of Physics and Astronomy, 
University of Georgia, Athens, GA 30602, USA}

\date{\today}

\begin{abstract}

Derivation of the Lorentz transformation without the use of Einstein's Second Postulate 
is provided along the lines of Ignatowsky, Terletskii, and others. This is a write-up of the 
lecture first delivered in PHYS 4202 E\&M class during the Spring semester of 2014 at the University of Georgia. 
The main motivation for pursuing this approach was to develop a better understanding of why the 
faster-than-light neutrino controversy (OPERA experiment, 2011) was much ado about nothing.

\end{abstract}


\maketitle


\section*{Special relativity as a theory of space and time}

All physical phenomena take place in space and time. 
The theory of space and time (in the absence of gravity) 
is called the Special Theory of Relativity. 

We do not get bogged down with the philosophical problems related to the 
concepts of space and time. We simply 
acknowledge the fact that in physics the notions of space and time are 
regarded as basic and cannot be reduced 
to something more elementary or fundamental. We therefore stick to 
pragmatic operational definitions: 

{\it Time is what clocks measure. Space is what measuring rods measure.}

In order to study and make conclusions about the properties of space and time 
we need an observer. 
A natural choice is an observer who moves freely (the one who is free from any 
external influences).
An observer is not a single person sitting at the origin of a rectangular coordinate grid. 
Rather, it is a bunch of friends (call it Team $K$) equipped with identical clocks distributed 
throughout the grid who record the events happening at their respective locations.  

\begin{figure}[!ht]
\includegraphics[angle=0,width=1.00\linewidth]{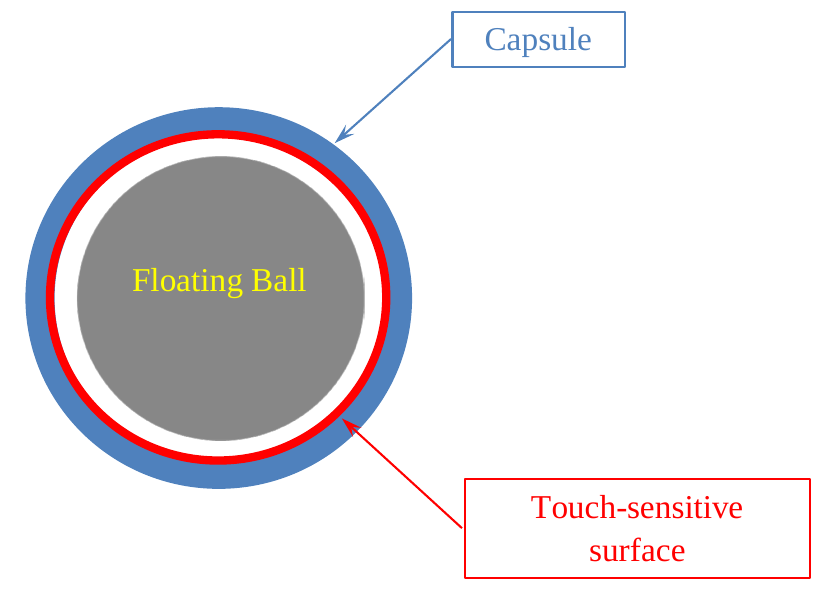}
\caption{ \label{fig:1} 
A floating-ball inertial detector. After \cite{MOORE2003}.
}
\end{figure}

How do we know that this bunch of friends is free from any external influences? 
We look around and make sure that nothing is pulling or pushing on any member 
of the bunch; no strings, no springs, no ropes are attached to them. 
An even better way is to use a collection of ``floating-ball detectors'' (Fig.\ \ref{fig:1})
distributed throughout the grid \cite{MOORE2003}. When detector balls are released, 
they should remain at rest inside their respective capsules. If any ball touches the 
touch-sensitive surface of the capsule, the frame is not inertial.

In the reference frame associated with a freely moving observer (our rectangular 
coordinate grid), Galileo's Law of Inertia is satisfied: a point mass, itself free from 
any external influences, moves with constant velocity. To be able to say 
what ``constant velocity'' really means, and thus to verify the law of inertia, we
need to be able to measure distances and time intervals between events happening 
at {\it different} grid locations.
 
\section*{Definition of clock synchronization}

It is pretty clear how to measure distances: the team simply uses its rectangular 
grid of rods.

It is also clear how to measure time intervals at a particular location: 
the team member situated at that location simply looks at his respective clock.
What's not so clear, however, is how the team measures time intervals between 
events that are {\it spacially} separated.

A confusion about measuring this kind of time intervals was going on for two hundred
years or so, until one day Einstein said: ``We need the notion of synchronized clocks!
Clock synchronization must be operationally defined.''

The idea that clock synchronization and, consequently, the notion of simultaneity of 
spacially separated events, has to be {\it defined} (and not assumed {\it apriori}) is 
the single most important idea of Einstein's, the heart of special relativity. Einstein 
proposed to use light pulses. The procedure then went like this: 

In frame $K$, consider two identical clocks equipped with light detectors, sitting 
some distance apart, at $A$ and $B$. Consider another clock equipped with a light 
emitter at location $C$ which is half way 
between $A$ and $B$ (we can verify that $C$ is indeed half way between $A$ 
and $B$ with the help of the grid 
of rods that had already been put in place when we constructed our frame $K$).  
Then, at some instant, emit two 
pulses from $C$ in opposite directions, and let those pulses arrive at $A$ and $B$. 
If the clocks at $A$ and $B$ show same time when the pulses arrive then the clocks 
there are synchronized, {\it by definition}.

The light pulses used in the synchronization procedure can be replaced with two 
identical balls initially sitting at $C$ and connected by a compressed spring. The 
spring is released (say, the thread holding the spring is cut in the middle), the balls 
fly off in opposite directions towards $A$ and $B$, respectively.   

How do we know that the balls are identical? Because Team $K$ made them in 
accordance with a specific manufacturing procedure.

How do we know that all clocks at $K$ are identical? Because Team $K$ made 
all of them in accordance with a specific manufacturing procedure.

How do we know that a tic-toc of any clock sitting in frame $K$ corresponds to 
1 second? Because Team $K$ called a tic-toc of a clock made in accordance with 
the manufacturing procedure ``a second''. 

Similarly, clocks in $K'$ are regarded as identical and tick-tocking at 1 second 
intervals because in that frame all of the clocks were made in accordance with 
the same manufacturing procedure.

Now, how do we know that the manufacturing procedures in $K$ and $K'$ are the same? 
(Say, how do we know that a Swiss shop in frame $K$ makes watches the same way as its 
counterpart in frame $K'$?) Hmm\dots. That's an interesting question to ponder about.

\vskip10pt 

When studying spacetime from the point of view of inertial frames of reference discussed above, 
people discovered the following.

\vskip10pt 

\begin{figure}[!ht]
\includegraphics[angle=0,width=1.00\linewidth]{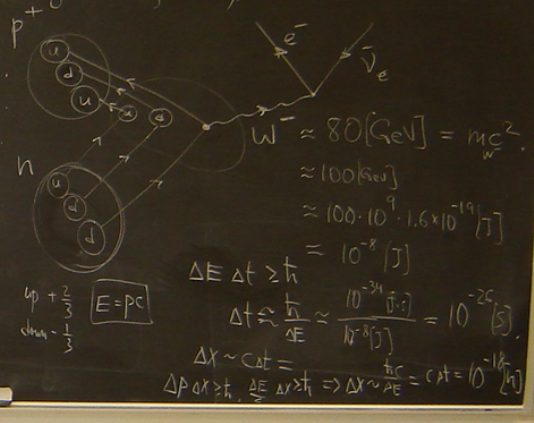}
\caption{ \label{fig:2} 
Neutron decay: Time is continuous down to $10^{-26}$ s.
}
\end{figure}

{\bf Properties of space and time:}
\begin{enumerate}
\item At least one inertial reference frame exists. 
(Geocentric is OK for crude experiments; geliocentric is better; the frame in which 
microwave background radiation is uniform is probably closest to ideal).
\item Space is uniform (translations; 3 parameters).
\item Space is isotropic (rotations; 3 parameters).
\item Time is uniform (translation; 1 parameter).
\item Space is continuous (down to $\sim 10^{-18}$ m).
\item Time is continuous (down to $\sim 10^{-26}$ s, Fig.\ \ref{fig:2}).
\item Space is Euclidean (apart from local distortions, which we ignore; 
cosmological observations put the limit at $\sim 10^{26}$ [m], the size of the visible 
Universe; this property is what makes rectangular grids of rods possible).
\item Relativity Principle (boosts; 3 parameters).
\end{enumerate} 

Einstein constructed his theory of relativity on the basis of (1) The Principle of Relativity 
(laws of nature are the same in all inertial reference frames), and (2) The Postulate of 
the Constancy of the Speed of Light (the speed of light measured by {\it any} inertial 
observer is independent of the state of motion of the emitting body). 
[NOTE: This is {\it not} the same as saying that the speed of light emitted {\it and} 
measured in $K$ is the same as the speed of light emitted {\it and} measured 
in $K'$. This latter type of constancy of the speed of light is already implied by 
the principle of relativity.]

Here we want to stick to mechanics and push the derivation of the coordinate transformation
as far as possible without the use of the highly counterintuitive Einstein's Second Postulate. 
The method that achieves this will be presented below and was originally due to 
Vladimir Ignatowsky \cite{Ignatowsky1911}.

\newpage

\section*{Step 1: Galileo's Law of Inertia for freely moving particles}

\noindent implies the linearity of the coordinate transformation between $K$ and $K'$
(see Fig.\ \ref{fig:1}),
\BEqA
x'&=&\a_{11}(v)x+\a_{12}(v)y+\a_{13}(v)z+\a_{14}(v)t,
\\
y'&=&\a_{21}(v)x+\a_{22}(v)y+\a_{23}(v)z+\a_{24}(v)t,
\\
z'&=&\a_{31}(v)x+\a_{32}(v)y+\a_{33}(v)z+\a_{34}(v)t,
\\
t'&=&\a_{41}(v)x+\a_{42}(v)y+\a_{43}(v)z+\a_{44}(v)t.
\EEqA
Here we assumed that the origins of the two coordinate systems coincide, that is
event $(0,0,0,0)$ in $K$ has coordinates $(0,0,0,0)$ in $K'$. 
\begin{figure}[!ht]
\includegraphics[angle=0,width=1.00\linewidth]{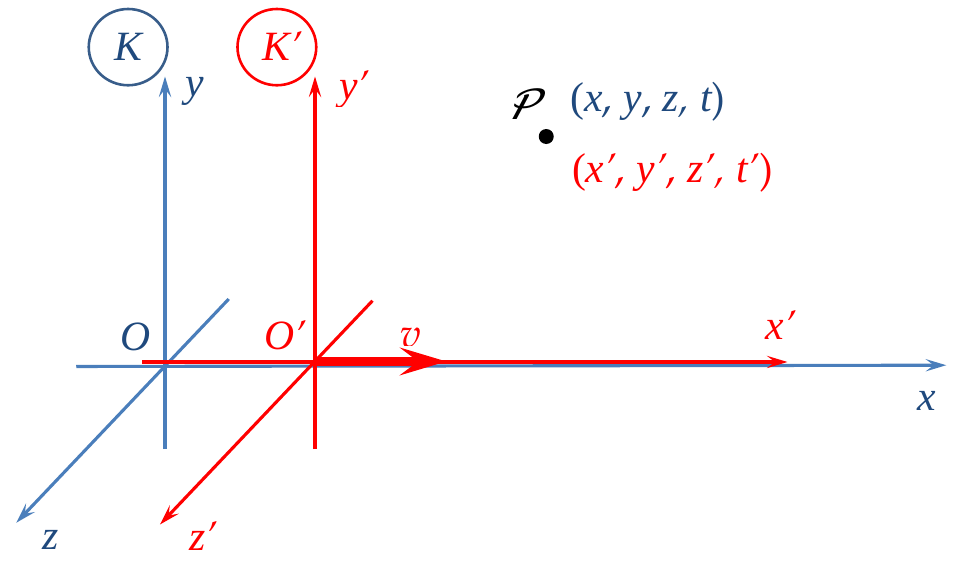}
\caption{ \label{fig:3} 
Two inertial reference frames (orthogonal grids of rods equipped 
with synchronized clocks) in relative motion along the $x$-axis.
}
\end{figure}

\section*{Step 2: The requirement that the 
$x'$-axis (line $y'=z'=0$) always coincides with the $x$-axis (line $y=z=0$) 
}

\noindent implies that $\a_{21}(v)=\a_{24}(v)=0$ and $\a_{31}(v)=\a_{34}(v)=0$, 
and thus $y'$ and $z'$ are independent of $x$ and $t$,
\BEqA
x'&=&\a_{11}(v)x+\a_{12}(v)y+\a_{13}(v)z+\a_{14}(v)t,
\\
y'&=&\a_{22}(v)y+\a_{23}(v)z,
\\
z'&=&\a_{32}(v)y+\a_{33}(v)z,
\\
t'&=&\a_{41}(v)x+\a_{42}(v)y+\a_{43}(v)z+\a_{44}(v)t.
\EEqA

\section*{Step 3: The requirement that the 
$x'y'$-plane ($z'=0$) always coincides with the $xy$-plane 
($z=0$) }

\noindent implies that $\a_{32}(v)=0$ and thus $z'$ is independent of $y$,
\BEqA
x'&=&\a_{11}(v)x+\a_{12}(v)y+\a_{13}(v)z+\a_{14}(v)t,
\\
y'&=&\a_{22}(v)y+\a_{23}(v)z,
\\
z'&=&\a_{33}(v)z,
\\
t'&=&\a_{41}(v)x+\a_{42}(v)y+\a_{43}(v)z+\a_{44}(v)t.
\EEqA

\section*{Step 4: Similarly, the requirement that the 
$z'x'$-plane ($y'=0$) always coincides with the $zx$-plane ($y=0$)}

\noindent implies that $\a_{23}(v)=0$,
and thus $y'$ is independent of $z$,
\BEqA
x'&=&\a_{11}(v)x+\a_{12}(v)y+\a_{13}(v)z+\a_{14}(v)t,
\\
y'&=&\a_{22}(v)y,
\\
z'&=&\a_{33}(v)z,
\\
t'&=&\a_{41}(v)x+\a_{42}(v)y+\a_{43}(v)z+\a_{44}(v)t.
\EEqA

\section*{Step 5: Isotropy of space}
 
\noindent also implies that $y'$ and $z'$ are physically equivalent,
so that $\a_{22}(v)=\a_{33}(v) \equiv k(v)$, and, thus,
\BEqA
x'&=&\a_{11}(v)x+\a_{12}(v)y+\a_{13}(v)z+\a_{14}(v)t,
\\
y'&=&k(v)y,
\\
z'&=&k(v)z,
\\
t'&=&\a_{41}(v)x+\a_{42}(v)y+\a_{43}(v)z+\a_{44}(v)t.
\EEqA

\section*{Step 6: Clock at $O'$ (the origin of frame $K'$) is moving 
with velocity $v$ along the $z$-axis as measured in $K$}

\noindent Correspondingly, we have:

\begin{enumerate}

\item In $K'$, $x'_{O'} = 0$. 

\item In $K$, 
\BEq
x_{O'}=vt, \quad y_{O'}=0, \quad z_{O'}=0.
\EEq
Since
\BEq
x'_{O'} = \a_{11}(v)x_{O'}+\a_{14}(v)t = 0, \quad \forall t,
\EEq
we have,
\BEq
(\a_{11}(v)v+\a_{14}(v))t = 0, \quad \forall t,
\EEq
or,
\BEq
\a_{14}(v)=- \a_{11}(v)v.
\EEq

\item Re-labeling $\a_{11}\equiv \a$,
this gives,
\BEqA
\label{eq:25}
x'&=&\a(v)(x - vt)+\a_{12}(v)y+\a_{13}(v)z,
\\
y'&=&k(v)y,
\\
z'&=&k(v)z,
\\
t'&=&\a_{41}(v)x+\a_{42}(v)y+\a_{43}(v)z+\a_{44}(v)t.
\EEqA

\item Since the entire $y'z'$-plane ($x'=0$) containing $O'$ (the origin of frame $K'$) is moving with velocity $v$ as measured in $K$, Eq.\ (\ref{eq:25}) should be true for any other clock belonging to the $y'z'$-plane. Isotropy of space then demands that 
$\a_{12}(v)=\a_{13}(v)=0$, $\a_{42}(v)=\a_{43}(v)=0$. After
re-labeling, $\a_{41}\equiv \d$ and $\a_{44}\equiv \g$, we get,
\BEqA
x'&=&\a(v)(x - vt),
\\
y'&=&k(v)y,
\\
z'&=&k(v)z,
\\
\label{eq:36}
t'&=&\d(v)x+\g(v)t.
\EEqA

\end{enumerate}

\section*{Step 7: Remarks}

\noindent {\bf NOTE:} The $\g$ just introduced will soon become the celebrated 
{\it gamma factor}.

\vskip10pt

\noindent {\bf IMPORTANT:} Eq.\ (\ref{eq:36}) indicates that it is possible to have two spatially separated events $A$ and $B$ 
that are simultaneous in frame $K$ and, yet, non-simultaneous in frame $K'$, that is
\BEq
\label{eq:9}
\Delta t_{AB} = 0, \; \Delta x_{AB} \neq 0: \; \Delta t'_{AB} = \d(v)\Delta x_{AB}\neq 0.
\EEq
This is not as obvious as might seem: for example, before Einstein it was assumed that whenever 
$\Delta t_{AB}$ is zero, $\Delta t'_{AB}$ must also be zero. So keeping $\d(v)$ in (\ref{eq:36}) 
is a significant departure from classical Newtonian
mechanics. 

\section*{Step 8: Inversion $\tilde{x} = -x$, $\tilde{y} = -y$, and 
$\tilde{x}' = -x'$, $\tilde{y}' = -y'$ }

\begin{figure}[!ht]
\includegraphics[angle=0,width=1.00\linewidth]{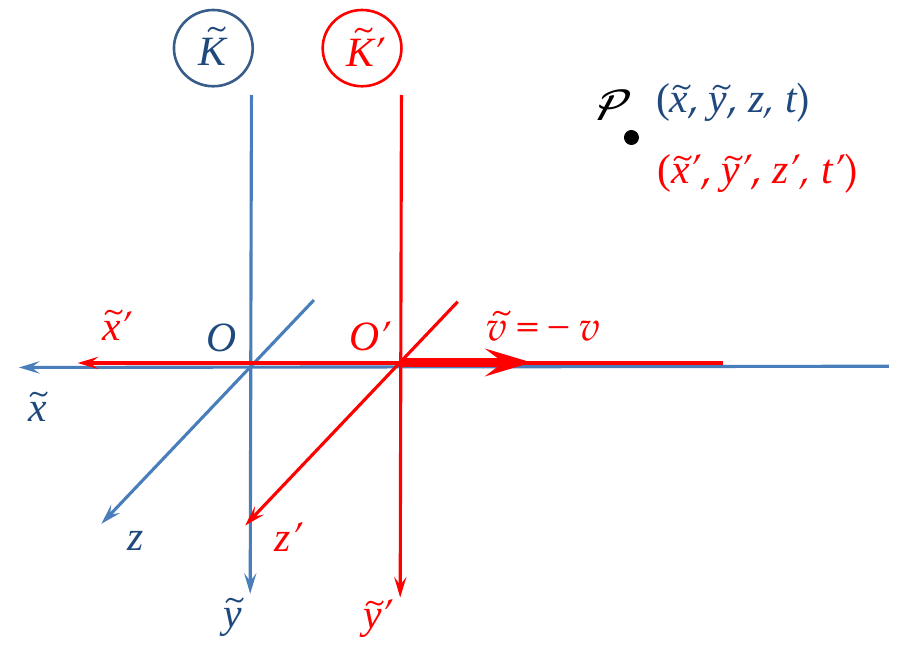}
\caption{ \label{fig:4} 
``Inverted'' frames in relative motion.
}
\end{figure}
 
\noindent  which is just a relabeling of coordinate marks, 
preserves the right-handedness of the coordinate 
systems and is physically equivalent to 
(inverted) frame $\tilde{K}'$ moving with velocity 
$\tilde{v} = -v$
relative to (inverted) frame $\tilde{K}$, so that
\BEqA
\tilde{x}'&=&\a(\tilde{v})(\tilde{x} - \tilde{v}t),
\\
\tilde{y}'&=&k(\tilde{v})\tilde{y},
\\
z'&=&k(\tilde{v})z,
\\
t'&=&\d(\tilde{v})\tilde{x}+\g(\tilde{v})t,
\EEqA
or,
\BEqA
-{x}'&=&\a(-v)(-{x} + vt),
\\
-{y}'&=&-k(-v){y},
\\
z'&=&k(-v)z,
\\
t'&=&-\d(-v){x}+\g(-v)t,
\EEqA
which gives
\BEqA
\a(-v)&=&\a(v),
\\
k(-v)&=&k(v),
\\
\d(-v)&=&-\d(v),
\\
\g(-v)&=&\g(v).
\EEqA

\section*{Step 9: Relativity principle and isotropy of space}
 
\noindent  tell us that 
the velocity of $K$ relative to ${K}'$, as measured by $K'$ using
primed coordinates $(x',t')$, is equal to $-v$. 

\vskip10pt

{\bf REMINDER:} the velocity of $K'$ relative to ${K}$,
as measured by $K$ using unprimed coordinates $(x,t)$, is $v$.

\vskip10pt

{\bf PROOF:} I justify the above claim by considering two local observers co-moving with $O$ and $O'$, respectively, and 
firing identical spring guns in opposite directions at the moment when they pass each other. If the ball shot in the $+x$ direction by $O$ 
stays next to $O'$ then, by the relativity principle and isotropy of space, the ball shot in the $-x'$ 
direction by $O'$ should stay next to $O$. 
This means that the velocity of $O$ relative to $O'$ as 
measured by $K'$ is {\it negative} of the velocity of $O'$ relative to $O$ as measured by $K$.

\vskip10pt

Thus, the inverse transformation is
\BEqA
{x}&=&\a(-v)({x}' + vt'),
\\
{y}&=&k(-v){y'},
\\
z&=&k(-v)z',
\\
t&=&\d(-v){x}'+\g(-v)t',
\EEqA
and since
\BEq
{y}=k(-v){y'}=k(-v)k(v)y=k^2(v)y,
\EEq
we get
\BEq
k(v) = \pm 1.
\EEq
Choosing $k(v)=+1$, which corresponds to parallel relative orientation of $y$ and $y'$ 
(as well as of $z$ and $z'$), gives, for the direct transformation,
\BEqA
x'&=&\a(v)(x - vt),
\\
y'&=&y,
\\
z'&=&z,
\\
t'&=&\d(v)x+\g(v)t,
\EEqA
and, for the inverse transformation,
\BEqA
{x}&=&\a(v)({x}' + vt'),
\\
{y}&=&{y'},
\\
z&=&z',
\\
t&=&-\d(v){x}'+\g(v)t'.
\EEqA

\section*{Step 10: Motion of $O$ (the origin of frame $K$)}
 
\begin{enumerate}

\item As seen from $K$: $x_{O}=0$.

\item As seen from $K'$: 
\begin{align}
x'_{O} &= \a(v)(x_{O}-vt_{O})=-v\a(v)t_{O},
\\
t'_{O}&= \d(v)x_{O}+\g(v)t_{O}=\g(v)t_{O}.
\end{align} 

\item But 
\BEq
x'_{O}= - v t'_{O},
\EEq 
which gives 
\BEq
- v \g(v)t_{O}=-v\a(v)t_{O}, 
\EEq
or,
\BEq
\a(v) = \g(v).
\EEq

\item
As a result,
\BEqA
x'&=&\g(v)(x - vt),
\\
y'&=&y,
\\
z'&=&z,
\\
t'&=&\d(v)x+\g(v)t,
\EEqA
and 
\BEqA
{x}&=&\g(v)({x}' + vt'),
\\
{y}&=&{y'},
\\
z&=&z',
\\
t&=&-\d(v){x}'+\g(v)t',
\EEqA
or, in matrix form,
\BEq
\begin{bmatrix}
x' \\ t'
\end{bmatrix}
=
\begin{bmatrix}
\g(v) & -v\g(v)  \\
\d(v) & \g(v)
\end{bmatrix}
\begin{bmatrix}
x \\ t
\end{bmatrix},
\EEq
and
\BEq
\begin{bmatrix}
x \\ t
\end{bmatrix}
=
\begin{bmatrix}
\g(v) & +v\g(v)  \\
-\d(v) & \g(v)
\end{bmatrix}
\begin{bmatrix}
x' \\ t'
\end{bmatrix}.
\EEq

\end{enumerate}

\section*{Step 11: The odd function $\d(v)$}
 
\noindent can be written as
\BEq
\d(v) = -vf(v^2)\g(v),
\EEq
since $\g(v)$ is even. 

\vskip10pt

{\bf NOTE:} The newly introduced function $f$ of $v^2$ will turn out to be a 
constant. Actually, one of the goals of the remaining steps of this derivation is to show that $f$ is a constant. It will later be identified with $1/c^2$.

\vskip10pt

\noindent Therefore,
\BEq
\begin{bmatrix}
x' \\ t'
\end{bmatrix}
=
\g
\begin{bmatrix}
1 & -v  \\
-vf & 1
\end{bmatrix}
\begin{bmatrix}
x \\ t
\end{bmatrix},
\EEq
and
\BEq
\begin{bmatrix}
x \\ t
\end{bmatrix}
=
\g
\begin{bmatrix}
1 & v  \\
vf & 1
\end{bmatrix}
\begin{bmatrix}
x' \\ t'
\end{bmatrix}.
\EEq

\section*{Step 12: Lorentz transformation followed by its inverse must give the identity transformation}
 
\noindent This seems physically reasonable. We have,
\BEqA
\begin{bmatrix}
x' \\ t'
\end{bmatrix}
&=&
\g
\begin{bmatrix}
1 & -v  \\
-vf & 1
\end{bmatrix}
\begin{bmatrix}
x \\ t
\end{bmatrix}
\nonumber \\
&=&
\g^2
\begin{bmatrix}
1 & -v  \\
-vf & 1
\end{bmatrix}
\begin{bmatrix}
1 & v  \\
vf & 1
\end{bmatrix}
\begin{bmatrix}
x' \\ t'
\end{bmatrix}
\nonumber \\
&=&
\g^2
\begin{bmatrix}
1-v^2f & 0  \\
0 & 1-v^2f
\end{bmatrix}
\begin{bmatrix}
x' \\ t'
\end{bmatrix},
\EEqA
and thus
\BEq
\g^2(1-v^2f)=1,
\EEq
from where
\BEq
\g = \pm \frac{1}{\sqrt{1-v^2f}}.
\EEq
To preserve the parallel orientation of the $x$ and $x'$ axes we have to choose the plus sign (as can be seen by taking the $v\rightarrow 0$ limit), so that
\BEq
\g = \frac{1}{\sqrt{1-v^2f}}.
\EEq
Thus,
\BEq
\begin{bmatrix}
x' \\ t'
\end{bmatrix}
=
\frac{1}{\sqrt{1-v^2f}}
\begin{bmatrix}
1 & -v  \\
-vf & 1
\end{bmatrix}
\begin{bmatrix}
x \\ t
\end{bmatrix},
\EEq
and
\BEq
\begin{bmatrix}
x \\ t
\end{bmatrix}
=
\frac{1}{\sqrt{1-v^2f}}
\begin{bmatrix}
1 & v  \\
vf & 1
\end{bmatrix}
\begin{bmatrix}
x' \\ t'
\end{bmatrix},
\EEq
where, we recall, $f = f(v^2)$.

\section*{Step 13: Two Lorentz transformations performed in succession is a Lorentz transformation}
 
This step is crucial for everything that we've been doing so far, for it shows that $f$ is a constant, which will be identified with $1/c^2$, where $c$ is Nature's limiting speed.

We have a sequence of two transformations: first from $(x,t)$ to $(x',t')$, then from
$(x',t')$ to $(x'',t'')$: 
\BEqA
\label{eq:100}
\begin{bmatrix}
x'' \\ t''
\end{bmatrix}
&=&
\g'
\begin{bmatrix}
1 & -v'  \\
-v'f' & 1
\end{bmatrix}
\begin{bmatrix}
x' \\ t'
\end{bmatrix}
\nonumber \\
&=&
\g'
\begin{bmatrix}
1 & -v'  \\
-v'f' & 1
\end{bmatrix}
\g
\begin{bmatrix}
1 & -v  \\
-vf & 1
\end{bmatrix}
\begin{bmatrix}
x \\ t
\end{bmatrix}
\nonumber \\
&=&
\g'\g
\begin{bmatrix}
1+vv'f & -(v+v')  \\
-(vf+v'f') & 1+vv'f'
\end{bmatrix}
\begin{bmatrix}
x \\ t
\end{bmatrix},
\EEqA
where $v$ is the velocity of $K'$ relative to $K$ (as measured in $K$ using the $(x,t)$ coordinates), 
and $v'$ is the velocity of $K''$ relative to $K'$ (as measured in $K'$ using the $(x',t')$ coordinates). But this could also be written as a single transformation from $(x,t)$ to $(x'',t'')$,
\BEqA
\label{eq:101}
\begin{bmatrix}
x'' \\ t''
\end{bmatrix}
&=&
\g''
\begin{bmatrix}
1 & -v''  \\
-v''f'' & 1
\end{bmatrix}
\begin{bmatrix}
x \\ t
\end{bmatrix},
\EEqA
with $v''$ being the velocity of $K''$ relative to $K$ (as measured in $K$ using the $(x,t)$ coordinates). 
This shows that the (1, 1) and (2, 2) elements of the transformation matrix (\ref{eq:100}) must be 
equal to each other and, thus,
\BEq
f=f',
\EEq
which means that $f$ is a constant that has units of inverse speed squared,
${\rm s}^2/{\rm m}^2$. 
{\bf Wow!!}

\section*{Step 14: Velocity addition formula (for reference frames)}
 
\noindent To derive the velocity addition formula
(along the $x$-axis) we use Eqs. (\ref{eq:100}) and (\ref{eq:101}) to get,
\BEqA
\g'\g (v+v') &=& \g'' v'',
\\
\g \g' (1+vv' f) &=& \g'',
\EEqA
which gives,
\BEq
\label{eq:102}
v''=\frac{v+v'}{1+vv'f}.
\EEq

\section*{Step 15: The universal constant $f$ cannot be negative }
 
\noindent because in that case the conclusions of relativistic dynamics 
would violate experimental observations. For example, the force law,
\BEq
\label{eq:forceEquation}
\frac{d}{dt}\frac{m{\bf v}_{p}}{\sqrt{1-v_{p}^2f}} = {\bf F},
\EEq
where ${\bf v}_p$ is the velocity of a particle, would get messed up \cite{TERLETSKII1968}. 
In particular, such law would violate the observation that
it requires an infinite amount of work (and, thus, energy) to accelerate a material
particle from rest to speeds approaching $3\times10^8$ m/s.
Incidentally, this {\it experimental} fact is what ``replaces'' Einstein's Second Postulate (that is, Michelson-Morley) in the present derivation. Otherwise, it would become ``easier'' to accelerate the particle, the faster it is moving. Thus, Eq.\ (\ref{eq:102})
is the limit to which our (actually, Ignatowski's) derivation can be pushed.

\vskip10pt

{\bf REMARK:} Relativistic dynamics has to be discussed separately, 
but maybe you can suggest 
a different reason for $f$ {\it not} to be negative? 
(See, e.\ g., \cite{BERZI1969} and \cite{MERMIN1984} for possible 
approaches.)

\section*{Step 16: Existence of the limiting speed}

\noindent Denoting
\BEq
f \equiv \frac{1}{c^2},
\EEq
we get the {\it velocity addition formula},
\BEq
\label{eq:103}
v''=\frac{v+v'}{1+\frac{vv'}{c^2}}.
\EEq
 If we start with $v'<c$ and attempt to take the limit $v'\rightarrow c$, we get
\BEq
\label{eq:104}
v''\rightarrow\frac{v+c}{1+\frac{vc}{c^2}}=c,
\EEq
which tells us that $c$ is the limiting speed that a material object can attain. 
[Notice that ``material'' here means  ``the one with which an inertial frame can be associated''. The photons do not fall into this category, as will be discussed shortly.]
The possibilities therefore are:
\begin{enumerate}
\item $c=+\infty$ (Newtonian mechanics; contradicts (\ref{eq:forceEquation}));
\item $c>0$ and finite (Special Relativity);
\item $c = 0$ (Contradics observations.)
\end{enumerate} 

So we stick with option 2.

What if an object were created to have $v'>c$ from the start (a so-called {\it tachyon}), 
like in the recent superluminal neutrino controversy (OPERA Experiment, 2011)? 
We'd get some strange results.

For example, if we take $v=c/2$ and $v' = 2c$, we get
\BEq
\label{eq:105}
v''=\frac{(c/2)+(2c)}{1+\frac{(c/2)(2c)}{c^2}}= \frac{5}{4}c,
\EEq
so in $K$ the object would move to the right at a smaller speed than relative to $K'$, while
$K'$ itself is moving to the right relative to $K$. 
Bizarre, but OK, the two speeds are measured by different 
observers, so maybe it's not a big deal \dots

\section*{Step 17: Lorentz transformation in standard form}

\noindent However, if we consider the resulting Lorentz transformation,
\BEq
\begin{bmatrix}
x' \\ t'
\end{bmatrix}
=
\frac{1}{\sqrt{1-\frac{v^2}{c^2}}}
\begin{bmatrix}
1 & -v  \\
-\frac{v}{c^2} & 1
\end{bmatrix}
\begin{bmatrix}
x \\ t
\end{bmatrix},
\EEq
or,
\BEqA
\label{eq:77}
t'&=&\frac{t-\frac{v}{c^2}x}{\sqrt{1-\frac{v^2}{c^2}}},
\\
\label{eq:78}
x'&=&\frac{x-vt}{\sqrt{1-\frac{v^2}{c^2}}},
\\
y'&=&y,
\\
z'&=&z,
\EEqA
we notice that in a reference frame $K'$ associated with hypothetical tachyons moving with $v>c$ 
relative to $K$ (imagine a whole fleet of them, forming a grid which makes up $K'$), the spacetime 
coordinates of any event would be {\it imaginary}.
In order for the spacetime measurements to give {\it real} values for
$(t',x',y',z')$, the reference frame $K'$ made of tachyons must be rejected. 

What about a reference frame made of photons?
In that case, coordinates would be infinite and should also be rejected. So a fleet of photons cannot form 
a ``legitimate'' reference frame. 
Nevertheless, we know that photons exist.
Similarly, tachyons may also exist and, like photons, 
(a) should be created instantaneously (that is, can't be 
created at rest, and then accelerated), and 
(b) should not be allowed to form a ``legitimate'' inertial reference 
frame. 

What about violation of causality?

\section*{Step 18: Violation of causality}

\noindent Indeed, the Lorentz transformation shows that tachyons violate causality. 
If we consider two events,
$A$ (tachyon creation) and $B$ (tachyon annihilation) with $t_B>t_A$ 
such that tachyon's speed, $v_p =  \frac{x_B-x_A}{t_B-t_A}>c$, is greater than $c$ 
as measured in $K$,
\begin{figure}[!ht]
\includegraphics[angle=0,width=1\linewidth]{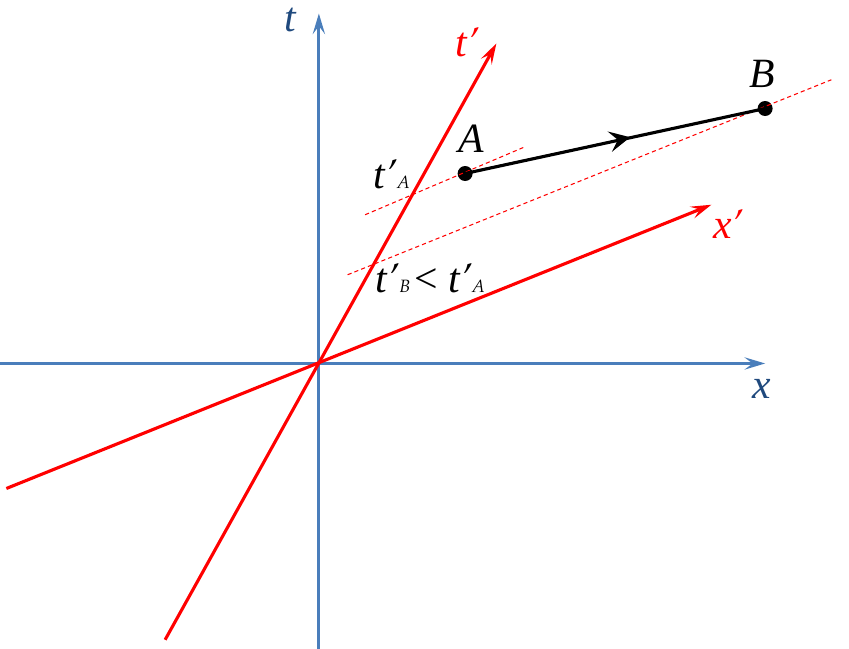}
\caption{ \label{fig:5} 
Violation of causality by a hypothetical tachyon.
}
\end{figure}
then in frame $K'$ moving with velocity $v<c$ relative to $K$ 
we'll have
from (\ref{eq:77}) and (\ref{eq:78}),
\BEqA
t'_{B}-t'_{A}&=&
\gamma
\left[
\left(t_{B}-t_{A}\right)
-\frac{v}{c^2}\left(x_{B}-x_{A}\right)
\right]
\nonumber \\
&=&
\gamma
\left[
1
-\frac{v}{c^2}\frac{x_{B}-x_{A}}{t_{B}-t_{A}}
\right]
\left(t_{B}-t_{A}\right)
\nonumber \\
&=&
\gamma
\left(
1
-\frac{vv_p}{c^2}
\right)
\left(t_{B}-t_{A}\right),
\\
x'_{B}-x'_{A}&=&
\gamma
\left[
\left(x_{B}-x_{A}\right)-v\left(t_{B}-t_{A}\right)
\right]
\nonumber \\
&=&
\gamma
\left[
1-v\frac{t_{B}-t_{A}}{x_{B}-x_{A}}
\right]\left(x_{B}-x_{A}\right)
\nonumber \\
&=&
\gamma
\left(
1-\frac{v}{v_p}
\right)\left(x_{B}-x_{A}\right),
\EEqA
where $\gamma = 1/\sqrt{1-v^2/c^2}$, which shows that it is possible to find $v<c$
such that  $t'_{B}-t'_{A} <0$; that is, in $K'$ event $B$ happens before event $A$. 
This seems to indicate that tachyons are impossible. 
However, causality is a consequence of the Second Law of Thermodynamics, which 
is a statistical law, applicable to macroscopic systems; it does not apply to processes
involving individual elementary particles. As a result, the existence of tachyons cannot be 
so easily ruled out.


\section*{Step 19: Speed of light is the limiting speed for material objects}

Finally, returning to Eq.\ (\ref{eq:104}), 
\[
v''\rightarrow\frac{v+c}{1+\frac{vc}{c^2}}=c,
\]
we see that if something moves with $c$ relative to $K'$,
it also moves with $c$ relative to any other frame $K''$. 
That is: the limiting speed is the same in all inertial reference frames. 
And there is no mentioning of any emitter, so 
we are recovering Einstein's Second Postulate.

Also, as follows from (\ref{eq:103}), 
\[
v''=\frac{v+v'}{1+\frac{vv'}{c^2}},
\]
$v'=c$ is the only speed that has this property (of being the same in all inertial frames). 
We know that light has this property (ala Michelson-Morley 
experiment), so the speed of light {\it is} the limiting speed for material objects. Since neutrinos have mass, they cannot move faster than light, and thus superluminal neutrinos are not possible.


\section*{Immediate consequences of the Lorentz transformation}

\subsection{Length contraction and relativity of simultaneity}


\includegraphics[angle=0,width=1.00\linewidth]{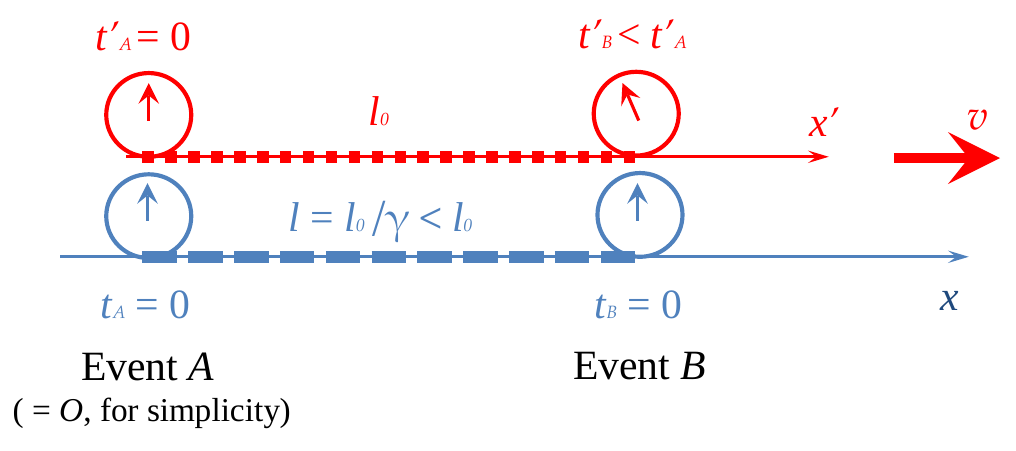}

\vskip5pt

Here we have a rod of (proper) length $\ell_0 \equiv  x_{B}'-x_{A}'>0$ 
sitting at rest in frame $K'$. 
Its speed relative to frame $K$ is $v$. The two events, $A$ and $B$, represent the 
meetings of the two clocks at the ends of the rod with the corresponding clocks in the $K$ frame at
$t_{A}=t_{B}$. We have 
from (\ref{eq:77}) and (\ref{eq:78}),
\BEqA
t'_{B}-t'_{A}&=&
\gamma
\left(-\frac{v}{c^2}\right)\left(x_{B}-x_{A}\right),
\\
x'_{B}-x'_{A}&=&
\gamma
\left(x_{B}-x_{A}\right),
\EEqA
or
\BEqA
\label{eq:85}
t'_{B}-t'_{A}&=&
\left(-\frac{v}{c^2}\right)\left(x'_{B}-x'_{A}\right),
\\
\label{eq:86}
x_{B}-x_{A}&=&\frac{x'_{B}-x'_{A}}{\gamma}.
\EEqA
Eq.\ (\ref{eq:85}) says that $t'_{B}-t'_{A}<0$, that is, the meeting events are not simultaneous in 
$K'$ ({\it relativity of simultaneity}). Eq.\ (\ref{eq:86}) says that the length of the rod, 
$\ell \equiv x_{B}-x_{A}$, as measured in $K$ is smaller than its proper length by the gamma factor,
\BEq
\ell = {\ell_0/\gamma},
\EEq
the phenomenon of {\it length contraction}.

\subsection{Time dilation}


\includegraphics[angle=0,width=1.00\linewidth]{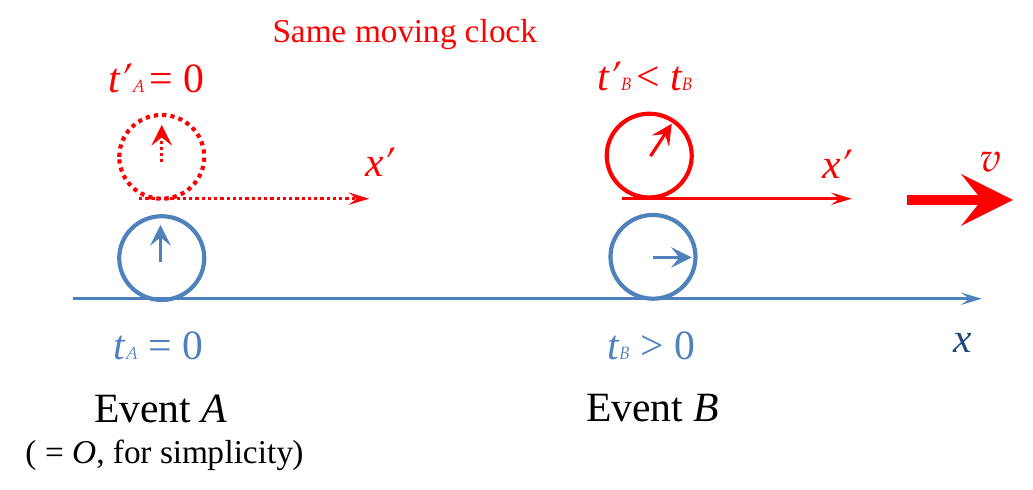}

This time a single clock belonging to $K'$ 
passes by two different clocks in $K$. 
The corresponding two events, 
$A$ and $B$, have
$x'_{A}=x'_{B}$, and are related to each other by
\BEqA
t'_{B}-t'_{A}&=&
\gamma
\left[
\left(t_{B}-t_{A}\right)-\frac{v}{c^2}\left(x_{B}-x_{A}\right)
\right]
\nonumber \\
&=& 
\gamma
\left[
1-\frac{v}{c^2}\frac{x_{B}-x_{A}}{t_{B}-t_{A}}
\right]\left(t_{B}-t_{A}\right)
\nonumber \\
&=& 
\gamma
\left(
1-\frac{v^2}{c^2}\right)\left(t_{B}-t_{A}\right)
\nonumber \\
&=& 
\frac{t_{B}-t_{A}}{\gamma}.
\EEqA
This means that upon arrival at $B$ the moving clock will read less time than
the $K$-clock sitting at that location. This phenomenon is called {\it time dilation} (moving clocks 
run slower).

\begin{acknowledgments}

I thank Todd Baker, Amara Katabarwa, and Loris Magnani for helpful discussions.

\end{acknowledgments}

\end{document}